\documentclass[sigconf,10pt]{acmart}

\usepackage{xspace}

\usepackage{balance}
\usepackage[capitalise]{cleveref} 
\crefname{section}{Sect.}{Sects.}
\crefname{figure}{Fig.}{Figs.}
\crefname{table}{Tab.}{Tabs.}
\crefname{equation}{Eq.}{Eqs.}

\usepackage{glossaries} 
\glsdisablehyper

\usepackage{siunitx} 
\usepackage{multirow} 
\usepackage{tabularx}
\usepackage{tikz}
\usepackage{bm}
\usepackage{pgfplots}
\pgfplotsset{compat=1.18}
\usepgfplotslibrary{fillbetween}
\usepgfplotslibrary{groupplots}

\usepackage[inline]{enumitem} 

\AtBeginDocument{%
  }
\copyrightyear{2026}
\acmYear{2026}
\setcopyright{cc}
\setcctype{by}
\acmConference[WiNTECH '26]{20th ACM Workshop on Wireless Network Testbeds, Experimental evaluation \& Characterization }{October 26--30, 2026}{Austin, TX, USA}
\acmBooktitle{20th ACM Workshop on Wireless Network Testbeds, Experimental evaluation \& Characterization (WiNTECH '26), October 26--30, 2026, Austin, TX, USA}
\acmDOI{10.1145/3831662.3844168}
\acmISBN{979-8-4007-2879-2/2026/10}

\crefname{section}{Sect.}{Sects.}
\crefname{figure}{Fig.}{Figs.}
\crefname{table}{Tab.}{Tabs.}
\crefname{equation}{Eq.}{Eqs.}
\newcommand{\on}{\mbox{ON}\xspace}
\newcommand{\off}{\mbox{OFF}\xspace}

\newcommand{\matlab}{\mbox{MATLAB}\xspace}

\newacronym[plural=AoAs,firstplural= angles of arrival (AoAs)]{AoA}{AoA}{angle of arrival}
\newacronym[plural=AoAs,firstplural= angles of departure (AoDs)]{AoD}{AoD}{angle of departure}
\newacronym[plural=AoAs,firstplural=angles of arrival (AoAs)]{aoa}{AoA}{angle-of-arrival}
\newacronym[plural=AoDs,firstplural=angles of departure (AoDs)]{aod}{AoD}{angle-of-departure}
\newacronym[plural=CFRs,firstplural= channel frequency responses (CFRs)]{CFR}{CFR}{channel frequency response}
\newacronym[plural=CIRs,firstplural= channel impulse responses (CIRs)]{CIR}{CIR}{channel impulse response}
\newacronym[plural=RNNs,firstplural= recurrent neural networks (RNNs)]{RNN}{RNN}{recurrent neural network}
\newacronym{$k$-NN}{$k$-NN}{k-nearest neighbors}
\newacronym{5g}{5G}{fifth generation}
\newacronym{6g}{6G}{sixth generation}
\newacronym{BS}{BS}{base station}
\newacronym{CDF}{CDF}{cumulative distribution function}
\newacronym{CSI}{CSI}{channel state information}
\newacronym{DET}{DET}{detection error tradeoff}
\newacronym{FA}{FA}{false alarm}
\newacronym{LRT}{LRT}{likelihood ratio test} 
\newacronym{LSTM}{LSTM}{long short-term memory}
\newacronym{LT}{LT}{likelihood test}
\newacronym{MAP}{MAP}{maximum a posteriori probability}
\newacronym{MD}{MD}{misdetection}
\newacronym{MIMO}{MIMO}{multiple-input multiple-output}
\newacronym{ML}{ML}{machine learning}
\newacronym{NMSE}{NMSE}{normalized mean-square-error}
\newacronym{PLA}{PLA}{physical layer authentication}
\newacronym{RIS}{RIS}{reconfigurable intelligent surface}
\newacronym{SNR}{SNR}{signal-to-noise ratio}
\newacronym{UE}{UE}{user equipment}
\newacronym{ULA}{ULA}{uniform linear array}
\newacronym{agc}{AGC}{automatic gain control}
\newacronym{ai}{AI}{artificial intelligence}
\newacronym{ans}{ANS}{advanced networking systems}
\newacronym{ap}{AP}{access point}
\newacronym{awgn}{AWGN}{additive white Gaussian noise}
\newacronym{bbm}{BBM}{black-box model}
\newacronym{cearc}{CEARC}{channel estimation with adaptive RIS configuration}
\newacronym{ce}{CE}{channel estimation}
\newacronym{cfo}{CFO}{carrier frequency offset}
\newacronym{cfr}{CFR}{channel frequency response}
\newacronym{cnn}{CNN}{convolutional neural network}
\newacronym{cots}{COTS}{commercial off-the-shelf}
\newacronym{csi}{CSI}{channel state information}
\newacronym{dft}{DFT}{discrete Fourier transform}
\newacronym{dl}{DL}{deep learning}
\newacronym{iq}{IQ}{in-phase and quadrature}
\newacronym{isac}{ISAC}{integrated sensing and communications}
\newacronym{lmb}{LMB}{linear model with bias}
\newacronym{lmmse}{LMMSE}{linear minimum mean squared error}
\newacronym{lm}{LM}{linear model}
\newacronym{los}{LOS}{line-of-sight}
\newacronym{ls}{LS}{least squares}
\newacronym{lts}{LTS}{long training symbol}
\newacronym{mbce}{MBCE}{MUSIC-based channel estimation}
\newacronym{mcs}{MCS}{modulation and coding scheme}
\newacronym{mimo}{MIMO}{multiple-input multiple-output}
\newacronym{miso}{MISO}{multiple-input single-output}
\newacronym{ml}{ML}{machine learning}
\newacronym{mmse}{MMSE}{minimum mean squared error}
\newacronym{mmwave}{mmWave}{millimeter-wave}
\newacronym{mrc}{MRC}{maximum ratio combining}
\newacronym{mse}{MSE}{mean squared error}
\newacronym{mu-mimo}{MU-MIMO}{multi-user MIMO}
\newacronym{music}{MUSIC}{multiple signal classification}
\newacronym{ndp}{NDP}{null data packet}
\newacronym{nlm}{NLM}{non-linear model}
\newacronym{nn}{NN}{neural network}
\newacronym{ofdm}{OFDM}{orthogonal frequency-division multiplexing}
\newacronym{omp}{OMP}{orthogonal matching pursuit}
\newacronym{parafac}{PARAFAC}{parallel factor}
\newacronym{pdf}{pdf}{probability density function}
\newacronym{pdr}{PDR}{packet delivery ratio}
\newacronym{phy}{PHY}{physical}
\newacronym{pls}{PLS}{physical-layer security}
\newacronym{pps}{PPS}{pulse-per-second}
\newacronym{raf}{RAF}{random forest}
\newacronym{relu}{ReLU}{rectified linear unit}
\newacronym{rf}{RF}{radio frequency}\newcommand{\wifi}{\mbox{Wi-Fi}\xspace}
\newacronym{ris}{RIS}{reflective intelligent surface}
\newacronym{rx}{RX}{receiver}
\newacronym{sdr}{SDR}{software-defined radio}
\newacronym{simo}{SIMO}{single-input multiple-output}
\newacronym{siso}{SISO}{single-input single-output}
\newacronym{snr}{SNR}{signal-to-noise ratio}
\newacronym{spdt}{SPDT}{single-pole double-throw}
\newacronym{ss}{SS}{spatial stream}
\newacronym{svd}{SVD}{singular value decomposition}
\newacronym{tx}{TX}{transmitter}
\newacronym{ue}{UE}{user equipment}
\newacronym{zf}{ZF}{zero-forcing}
\newacronym{vna}{VNA}{vector network analyzer}
\newacronym{fr4}{FR-4}{flame retardant 4}
\newacronym{pcb}{PCB}{printed circuit board}
\newacronym{sfo}{SFO}{sampling frequency offset}
\newacronym{pll}{PLL}{phase-locked loop}

\let\ac\gls
\let\acp\glspl

\begin{document}

\title[BRISC: A Dataset of RIS Channel Measurements at \SI{5}{\giga\hertz}]{BRISC: A Dataset of Channel Measurements at \SI{5}{\giga\hertz} With a Reconfigurable Intelligent Surface}

\author{Mattia Piana}
\orcid{0009-0009-2818-4861}
\affiliation{%
  \institution{University of Padova}
  \city{Padova}
  \country{Italy}}
\email{mattia.piana@phd.unipd.it}

\author{Giovanni Angelo Alghisi}
\authornote{Also with CNIT, National Inter-University Consortium for Telecommunications.}
\orcid{0009-0002-1760-6263}
\affiliation{%
  \institution{University of Brescia \& CNIT}
  \city{Brescia}
  \country{Italy}}
\email{giovanni.alghisi@unibs.it}

\author{Anna Valeria Guglielmi}
\orcid{0009-0001-3737-8168}
\affiliation{%
  \institution{University of Padova}
  \city{Padova}
  \country{Italy}}
\email{annavaleria.guglielmi@unipd.it}

\author{Giovanni Perin}
\authornotemark[1]
\orcid{0000-0002-7333-3004}
\affiliation{%
  \institution{University of Brescia \& CNIT}
  \city{Brescia}
  \country{Italy}}
\email{giovanni.perin@unibs.it}

\author{Francesco Gringoli}
\authornotemark[1]
\orcid{0000-0003-2621-582X}
\affiliation{%
  \institution{University of Brescia \& CNIT}
  \city{Brescia}
  \country{Italy}}
\email{francesco.gringoli@unibs.it}

\author{Stefano Tomasin}
\orcid{0000-0003-3253-6793}
\affiliation{%
  \institution{University of Padova}
  \city{Padova}
  \country{Italy}}
\email{stefano.tomasin@unipd.it}

\renewcommand{\shortauthors}{M.~Piana et al.}

\begin{abstract}
We introduce the broadband \ac{RIS} channel (BRISC) dataset. The dataset comprises measurements of \ac{CSI} collected at \SI{5.530}{\giga\hertz} using a 256-element \ac{RIS} with binary states. In the measurement campaign, the transmitter and receiver are two \acp{sdr}, phase-synchronized via an OctoClock, where the transmitter (receiver) is equipped with one (two) antenna(s). To manage complexity, the \ac{RIS} elements are grouped into blocks of different sizes, where all elements within a block share the same state.
\acp{CSI} have been captured for multiple a) transmitter positions (and fixed receiver location), b) \ac{RIS} element-grouping granularity, and c) \ac{RIS} configurations. Furthermore, we calibrated the parameters of state-of-the-art \ac{RIS} channel models to fit the measured \ac{CSI}. With approximately 10\,000 configurations explored per transmitting position, BRISC serves as a robust benchmark in communication applications.
\end{abstract}


\maketitle

\glsresetall

\section{Introduction}
Research on next-generation wireless systems is moving beyond the traditional paradigm of passive propagation environments.
As \ac{5g} is being deployed, research increasingly targets \ac{6g}, where extreme flexibility, reliability, and security are expected to play a central role.
In this context, \acp{RIS} have emerged as a promising solution to actively shape the wireless channel by dynamically controlling the electromagnetic response of the environment~\cite{direnzo2020sre, Wu2018}.

A \ac{RIS} consists of a large number of sub-wavelength reflecting elements whose electromagnetic response can be electronically tuned. By configuring these elements, a \ac{RIS} controls the phase and amplitude of the reflected signal~\cite{liuReconfigurable2021}, making impinging waves interfere constructively or destructively at desired locations, thereby enabling programmable beamforming and spatial filtering~\cite{Wu2018}.
In challenging indoor environments affected by multipath, shadowing, and weak line-of-sight, RISs can improve coverage and link quality~\cite{liuReconfigurable2021}.

\ac{RIS}-assisted communications also support emerging services such as localization, sensing, and \ac{pls}. In particular, a \ac{RIS} can enhance legitimate links while suppressing signal leakage and mitigating jamming or interference, thus improving confidentiality and resilience~\mbox{\cite{9198898, Niu2021, Niu2023}}.

Whether the goal is to design beamforming schemes or to develop \ac{pls} solutions, studying these techniques and validating the associated \ac{RIS}-assisted channel models require experimental data. So far, most works rely on simulations based on theoretical or ray-tracing channel models~\cite{masood2023inductive, Wu2018, ElMossallamy2021, Niu2023}, which are useful for initial algorithm design but can capture neither the behavior of a real \ac{RIS} prototype nor the hardware non-idealities of a practical deployment~\cite{heinrichs2023open-source-ris}.

Experimental validation is therefore essential, yet only a few datasets of communication channels through \acp{RIS} are  available~\cite{comprehensiveTewes2023, rossanese2024, dataSimulated}.
In particular, to the best of our knowledge, no publicly available dataset characterizes the \ac{RIS} behavior and its associated channel across a broad range of \ac{RIS} configurations under realistic and standard-compliant conditions, such as wideband \ac{ofdm} transmissions.
This gap hinders the reproducible study and fair comparison of the many techniques envisioned for \ac{RIS}-enabled systems.

This paper introduces the broadband \ac{RIS} channel (BRISC) dataset, with the following main contributions:
\begin{itemize}
    \item We release BRISC~\cite{zenodoBrisc}, a publicly available dataset of wideband \ac{CSI} measurements, i.e., the per-subcarrier frequency response of the wireless channel between transmitter and receiver. The dataset was collected with a $16\times16$ binary-state \ac{RIS} operating at \SI{5.530}{\giga\hertz} over an \SI{80}{\mega\hertz} IEEE~802.11\allowbreak{}ac \ac{ofdm} channel, using phase-synchronized \acp{sdr}, nine transmitter positions, and two receive antennas, for a total of approximately \num{10000} \ac{RIS} configurations per position.
    \item We explore the \ac{RIS} configuration space through a structured strategy, from coarse macro-block groupings to fully random element-wise states, at different levels of control granularity.
    \item We calibrate and benchmark linear and data-driven \ac{RIS} channel models against the measured data, revealing a configuration-independent bias term missed by standard formulations, and characterize the accuracy/training trade-offs across model families.
    \item We release the processing and analysis scripts used in this work~\cite{gitHubBrisc}, to support reproducibility and further research on \ac{RIS}-assisted communications.
\end{itemize}

\section{Related work}
As mentioned in the Introduction, the limited amount of publicly available data on \ac{CSI} with \acp{RIS} represents a major obstacle to reproducible research, model validation, and fair comparison of signal processing, learning, and security techniques designed for \ac{RIS}-assisted communications.

For instance, \cite{dataSimulated} releases a publicly available, but entirely simulated, dataset to support the validation of a deep-learning-based channel estimation method for \ac{RIS}-aided systems. The dataset is generated through simulation of a massive \ac{MIMO} uplink at \SI{28}{\giga\hertz}, considering an $8\times8$ base station array, a $16\times16$ \ac{RIS}, and six users, without any experimental validation on real hardware. In contrast, our work considers a different, experimentally-driven scenario, based on real wideband \ac{ofdm} transmissions with a physical \ac{RIS} prototype operating at \SI{5.530}{\giga\hertz}.

In~\cite{rossanese2024}, the authors publicly released two datasets of scalar received-power measurements, collected using a custom $10\times10$-element \ac{RIS} in an anechoic chamber.
The first sweeps a predefined codebook of beam-steering configurations, designed under a purely line-of-sight channel model, while a turntable jointly rotates the \ac{RIS} and transmitter with respect to a fixed receiver.
The second keeps the transmitter-\ac{RIS}-receiver geometry fixed and instead uses an absorption-mode feature to selectively deactivate elements, thereby measuring the beamforming gain achieved by square sub-arrays of increasing size.
However, these datasets rely on scalar received-power measurements rather than wideband \ac{CSI}, and the explored configurations are constrained to a codebook derived from an assumed channel model rather than covering the \ac{RIS} configuration space broadly.

In~\cite{comprehensiveTewes2023}, the authors employed the same \ac{RIS} we used in this work and publicly released a dataset of $S_{21}$ measurements, the scattering parameter quantifying the amplitude and phase of the signal from the transmit to the receive port, obtained with a \ac{vna} over the \SI{5}{}--\SI{6}{\giga\hertz} band (\num{201} points, \SI{5}{\mega\hertz} step).
For each geometric scenario, the \ac{RIS} switching state is obtained via iterative search algorithms (greedy and single-element optimization) that converge to a single configuration maximizing or minimizing the received power, so that only a handful of near-optimal states are captured per scenario.
While this constitutes a valuable, publicly available characterization of the achievable gain of the prototype, it does not support the study of \ac{RIS} channel models, which instead requires observing the channel response across a broad and representative set of configurations.
Our dataset addresses this need: the \ac{CSI} is estimated from real IEEE~802.11ac VHT frames, with a per-subcarrier resolution of \SI{312.5}{\kilo\hertz}, sixteen times finer than the step used in~\cite{comprehensiveTewes2023}.
The \SI{80}{\mega\hertz} bandwidth we considered is narrower than the \SI{1}{\giga\hertz} band explored in~\cite{comprehensiveTewes2023}, but representative of a real \ac{ofdm} system such as Wi-Fi, and it is centered at the frequency at which the \ac{RIS} was designed to provide a \ang{180} phase shift between its ON/OFF states, i.e., where its impact on the channel frequency response is strongest.
Moreover, we densely sample the \ac{RIS} configuration space, from coarse macro-blocks to fully random element-wise states (\num{10000} configurations per transmitter position), enabling the calibration and validation of data-driven \ac{RIS} channel models, as demonstrated in Section~\ref{sec:results}.



\section{Experimental Setup and Data Acquisition}
\label{sec:measurements}

\begin{figure}
    \centering
    \includegraphics[height=0.9\linewidth]{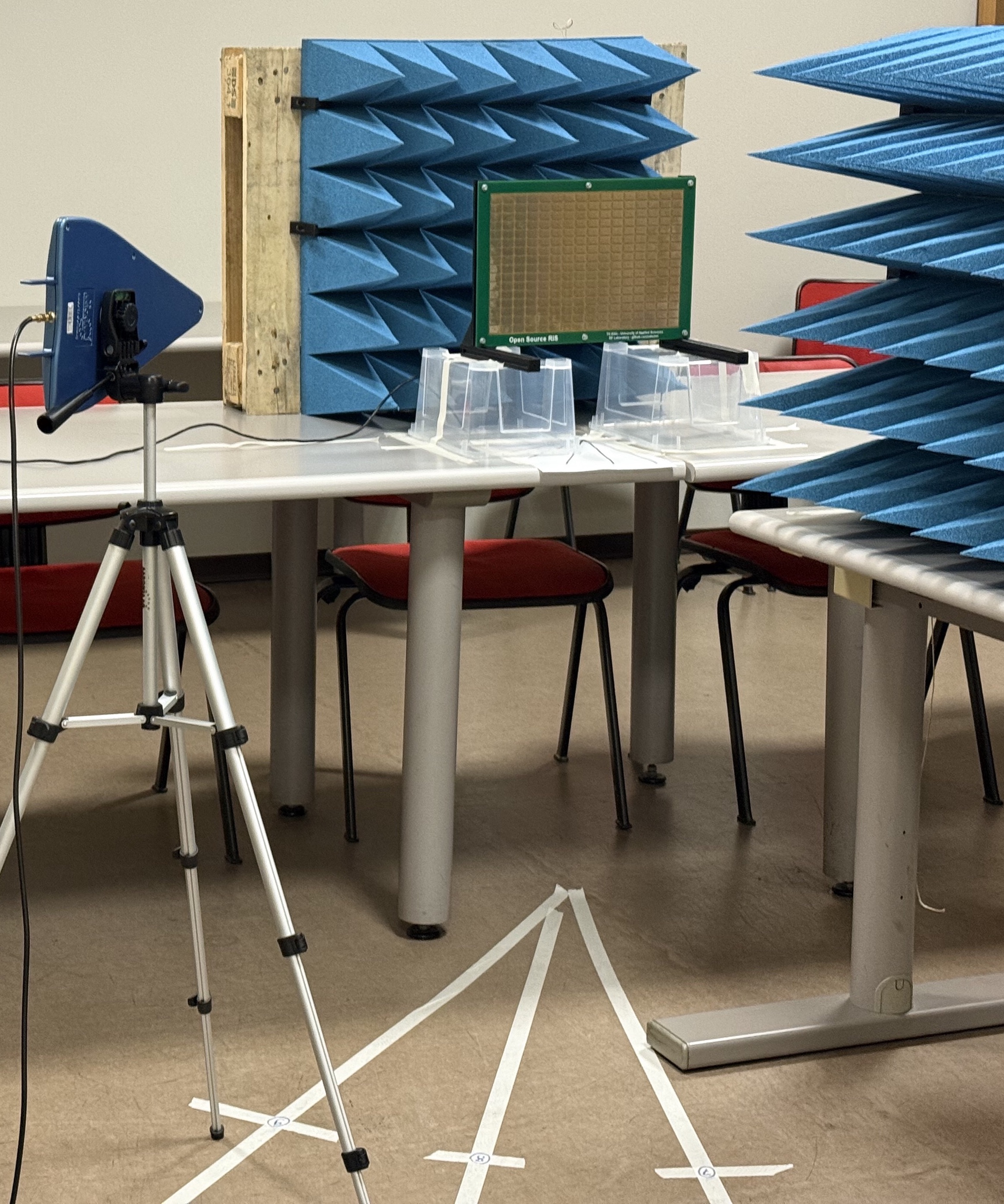}
    \caption{Photograph of the experimental testbed used for the measurement campaign, highlighting the transmitting antenna and the $16\times16$ \ac{RIS} prototype.}
    \Description{Photograph of the experimental testbed, showing a 16x16 Reconfigurable Intelligent Surface (RIS) prototype mounted on a support structure, the transmitting antenna, and anechoic absorbing panels placed in the background to reduce unwanted reflections.}
    \label{fig:photo-setup}
\end{figure}

This section presents the experimental testbed and the measurement campaign underlying the BRISC dataset. We detail the hardware setup, the data acquisition procedure, and the strategy adopted to systematically collect raw \ac{iq} recordings and \acp{CSI} under different \ac{RIS} configurations and \ac{tx} locations. 

\cref{fig:photo-setup} shows a picture of the experimental setup, including the \ac{RIS}, electromagnetic absorbing panels, and a log-periodic antenna for transmission. One absorber is positioned between the \ac{tx} and \ac{rx} to block the \ac{los} between the two, thus allowing the \ac{RIS} to have a stronger impact on the measured channel. Another absorber is positioned behind the \ac{RIS} to block wall reflections, thereby isolating the signal reflected by the \ac{RIS}.

\begin{figure}
    \centering
    \includegraphics[height=0.975\linewidth]{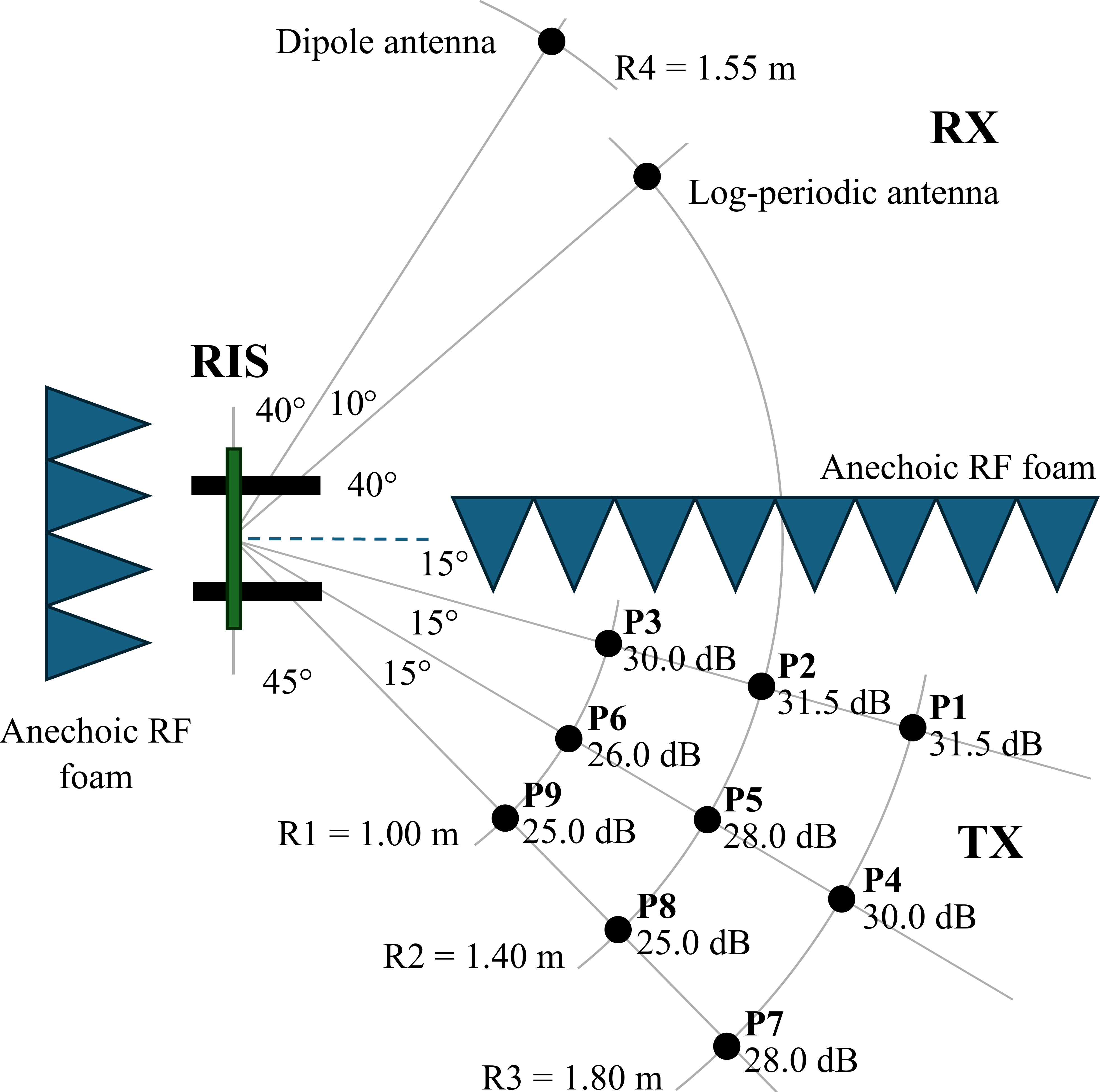}
    \caption{Measurement setup and geometry of the experimental scenario.}
    \Description{Diagram of the measurement setup showing a Reconfigurable Intelligent Surface (RIS) and the spatial arrangement of transmitter (TX) and receiver (RX) positions. Nine TX positions are arranged on a grid defined by three arcs of increasing radius centered on the RIS (R1 = 1 m, R2 = 1.4 m, R3 = 1.8 m) and three angular directions at 15, 30, and 45 degrees from the RIS normal; each TX position is labeled with its corresponding fixed TX gain. Two RX positions are shown at different distances and angles from the RIS: one for a log-periodic antenna at radius R2 and 40 degrees from the RIS normal, and one for a omnidirectional antenna at radius R4 = 1.55 m and 50 degrees from the RIS normal. Anechoic absorbing panels are placed behind the RIS and between the TX and RX regions to reduce unwanted reflections.}
    \label{fig:setup-scheme}
\end{figure}

During the experiments, the \ac{RIS} and the \ac{rx} remain fixed, while the \ac{tx} is placed at nine predefined positions. A top view of the overall setup geometry is illustrated in \cref{fig:setup-scheme}. The multiple \ac{tx} positions refer to the different locations of the single log-periodic transmitting antenna. At the \ac{rx}, two antennas (a log-periodic and a dipole) are simultaneously deployed and remain fixed in position throughout the entire measurement campaign.
Both the \ac{tx} and \ac{rx} log-periodic antennas are always pointed at the \ac{RIS}, so that the intended \ac{tx}--\ac{RIS}--\ac{rx} path is consistently illuminated.
For each \ac{tx} location, a suitable internal \ac{sdr} transmit gain was selected to adjust the signal to the range of the receiving \ac{sdr}, as reported in the figure.

\subsection{RIS Prototype}
\label{subsec:ris}

The \ac{RIS} employed in this study is described in \cite{heinrichs2023open-source-ris} and consists of a $16\times 16$ array of unit cells manufactured on a low-cost \ac{fr4} \ac{pcb}, for a total of 256 elements. The surface measures \SI{400}{\milli\meter} in width and \SI{320}{\milli\meter} in height. A dedicated control board is mounted on the back of the panel and connected to a host computer via USB, enabling real-time reconfiguration during the measurement campaign. For mechanical stability, the surface is fixed to an acrylic support using stand-offs and nylon screws.

Each unit cell follows a three-layer \ac{pcb} architecture with a plated through-hole (via) connecting the front radiating patch to the control circuitry on the backside, similar to a pin-fed patch antenna. The backside circuit integrates a CMOS-based \ac{spdt} \ac{rf} swit\-ch that connects the feed of the patch to one of two terminations, namely an open circuit or a short circuit, with low power consumption. The two terminations impose different boundary conditions on the patch, so the incident wave is re-radiated with two distinct complex reflection coefficients, providing 1-bit control of each element. The two coefficients differ in phase and, since the \ac{fr4} substrate and the switch are lossy and dissipate a different amount of power in the two states, also in amplitude.
The electromagnetic response of the surface has been experimentally characterized in~\cite{heinrichs2023open-source-ris}, where the complete hardware design and measurement campaign are detailed.
The device has been designed to maximize the phase variation across the \wifi frequency band from \SIrange{5.150}{5.875}{\giga\hertz}.


\subsection{Instrumentation Hardware and Experimental Setup}
\label{subsec:hw}
The acquisition system relies on two identical \acp{sdr}, serving as \ac{tx} and \ac{rx}, and specifically two Ettus USRP X310 platforms\footnote{\url{https://kb.ettus.com/X300/X310}} equipped with UBX-160 daughterboards.\footnote{\url{https://kb.ettus.com/UBX}}
The two \acp{sdr} are synchronized using an external OctoClock,%
\footnote{\url{https://kb.ettus.com/OctoClock_CDA-2990}} %
providing a common \SI{10}{\mega\hertz} reference clock and \ac{pps} signal.
This configuration ensures phase coherence between the \ac{tx} and the \ac{rx}, avoiding \ac{cfo} and \ac{sfo}, as well as making the phase offset introduced by the \ac{pll} constant over the whole acquisition, rather than random from frame to frame.
This is essential for reliable \ac{CSI} acquisition and phase-sensitive measurements.

To reduce uncontrolled reflections in the indoor environment, the \ac{tx} employs a highly directive log-periodic antenna (HyperLOG~60100).\footnote{\url{https://aaronia.com/en/log-periodic-antenna-hyperlog7060}}
At the \ac{rx} side, two antennas are used: a directive log-periodic antenna (HyperLOG~30100)\footnote{\url{https://aaronia.com/en/breitbandantenne-hyperlog30100}} and a dipole antenna (Ettus VERT2450).\footnote{\url{https://www.ettus.com/all-products/vert2450/}}
The log-periodic antenna provides spatial filtering and emphasizes the intended \ac{tx}-\ac{RIS}-\ac{rx} path, whereas the dipole antenna, due to its broader radiation pattern, captures a richer multipath structure.

To obtain meaningful \ac{CSI} measurements and accurately study the channel variations induced by the \ac{RIS}, the gain of the acquisition chain must remain fixed for each transmitter position, without \ac{agc}. Otherwise, variations introduced by the \ac{agc} itself would be indistinguishable from those caused by changes in the propagation channel, preventing a reliable interpretation of the measured \ac{CSI}.
For this reason, the receive chain operates with \ac{agc} disabled and a fixed internal gain of \SI{35}{\decibel}.

On the transmit side, an external linear power amplifier is used to increase the transmitted signal level. The amplifier operates in the band of  \SIrange{5}{5.8}{\giga\hertz}, with a nominal gain of \SI{25}{\decibel}, and a maximum output power of \SI{2}{\watt}.

Because channel attenuation varies across \ac{tx} locations, the received signal power changes accordingly. 
To fully exploit the \ac{sdr} dynamic range while avoiding receiver saturation, the internal transmit gain of the \ac{sdr} is statically adjusted for each \ac{tx} position, as indicated in \cref{fig:setup-scheme}. 
The gain tuning is performed considering the directive log-periodic antenna at the \ac{rx}, which provides higher antenna gain and therefore represents the worst-case condition in terms of potential receiver saturation. 
Once selected, the transmit gain remains fixed while sweeping over all the \ac{RIS} configurations at that position and is reconfigured only when the transmitter is moved to a different location.
We also verified that, for all \ac{tx} positions, the selected transmit gains keep the \ac{sdr} output power below the level that would drive the external amplifier into compression, so that, according to its datasheet, the amplifier operates in its linear region.


\section{Measurement and RIS Switching Procedure}
\label{subsec:measurement-procedure}
The measurement campaign is conducted using IEEE~802.11ac VHT frames over an \SI{80}{\mega\hertz} bandwidth centered at a carrier frequency of \SI{5.530}{\giga\hertz}. This frequency is selected as it corresponds to the operating point at which the \ac{RIS} introduces the largest phase variation~\cite{heinrichs2023open-source-ris}.

While the receiver continuously captures and stores complex baseband \ac{iq} samples for offline processing in \matlab, the transmitter operation and \ac{RIS} reconfiguration follow a structured sequence of recurring phases:

\begin{enumerate}    
    \item \emph{\ac{RIS} Configuration and Settling:}\label{phase:ris-config}  
    The \ac{RIS} is programmed with a given configuration pattern by setting the binary state of its unit cells. After applying the configuration, a settling time of approximately \SI{0.3}{\second} is enforced before transmission, giving the \ac{RIS} enough time to change its configuration.

    \item \emph{Frame Generation and Payload Tagging:}\label{phase:frame-generation}  
    \wifi frames are generated in \matlab, each one embedding an identifier corresponding to the active \ac{RIS} configuration, written at a fixed payload offset. This enables an unambiguous association between received frames and \ac{RIS} states during post-processing.

    \item \emph{Transmission Window:}\label{phase:tx-on}  
    The \ac{tx} streams the generated waveform continuously for approximately \SI{0.6}{\second}, while keeping the \ac{RIS} configuration fixed. During this interval, the \ac{rx} collects approximately 60~frames (thus, 60~\ac{CSI} samples per receiving antenna).

    \item \emph{Silence Window and Repetition:}\label{phase:tx-off}  
    After the transmission window, the \ac{tx} is configured to transmit zeros (i.e., remain effectively silent) for approximately \SI{0.3}{\second}. This silence interval facilitates temporal separation between consecutive \ac{RIS} configurations and simplifies the segmentation and parsing of the received data during offline processing. Upon completion of this interval, the procedure resumes from Phase~\ref{phase:ris-config}.
\end{enumerate}
The above procedure is repeated for all considered \ac{RIS} configurations and for nine different \ac{tx} positions, while keeping the \ac{rx} and \ac{RIS} locations fixed, as illustrated in~\cref{fig:setup-scheme}.

\subsection{Exploring the RIS Configuration Space}
\label{subsec:ris-config}
Given the extremely large configuration space of the $256$-element binary \ac{RIS}, we adopt a structured exploration strategy to balance analytical tractability and measurement diversity within a single measurement campaign.

The first $16$ configurations are obtained by partitioning the surface into four square macro-blocks of $8\times 8$ elements. All elements within the same macro-block share the same state (\on\ or \off), so that the \ac{RIS} can be virtually regarded as a four-element surface. This reduced representation enables a tractable analysis of channel behavior and facilitates the validation of simple linear models. In this setting, all possible $2^4 = 16$ combinations of the macro-block states are exhaustively explored.

Subsequently, the surface is divided into nine macro-blocks arranged in a $3\times3$ grid. The first two macro-rows and macro-columns consist of $5$ elements each, while the third macro-row and macro-column consist of $6$ elements each. Also in this case, all $2^9 = 512$ possible combinations of the macro-block states are measured. The remaining configurations are generated by independently assigning the \on/\off\ state to each element. Overall, this progressive strategy, from coarse macro-block control to fully random element-wise configurations, results in 10\,000 distinct \ac{RIS} configurations explored for each \ac{tx} position.

The main experimental parameters are reported in \cref{tab:experimental-parameters}.

\begin{table}[t]
    \centering
    \caption{Setup and measurement parameters}
    \label{tab:experimental-parameters}
    \small
    \setlength{\tabcolsep}{4pt}
    \begin{tabularx}{\columnwidth}{@{}l >{\raggedright\arraybackslash}X@{}}
        \toprule
        \textbf{Experimental detail}        & \textbf{Description} \\
        \midrule
        Carrier frequency                   & \SI{5.530}{\giga\hertz} \\
        Signal bandwidth                    & \SI{80}{\mega\hertz} (IEEE 802.11ac VHT) \\
        Valid subcarriers                   & 242 \\
        Recorded signals                    & Complex baseband \ac{iq} samples \\
        Extracted metric                    & \ac{CSI} \\
        \ac{RIS} elements                   & 256 (16$\times$16 array) \\
        \ac{RIS} control                    & 1-bit per element (binary states) \\
        \ac{RIS} operating band             & \SIrange{5.150}{5.875}{\giga\hertz} \\
        \ac{RIS} size                       & \SI{400}{\milli\meter} $\times$ \SI{320}{\milli\meter} \\
        Unit-cell architecture              & 3-layer PCB with CMOS \ac{spdt} switch \\
        \ac{sdr} platforms                  & 2$\times$ Ettus USRP X310 with UBX-160 daughterboards \\
        \ac{sdr} synchronization            & External OctoClock (\SI{10}{\mega\hertz} + \ac{pps}) \\
        \ac{tx} antenna                     & Log-periodic (HyperLOG 60100) \\
        \ac{rx} antennas                    & Log-periodic (HyperLOG 30100) and dipole (VERT2450) \\
        \ac{rx} \ac{sdr} gain               & \SI{35}{\decibel} (fixed, \ac{agc} off) \\
        \ac{tx} external amplifier          & \SI{25}{\decibel} gain, \SI{2}{\watt} max output \\
        \ac{tx} positions                   & 9 \\
        \ac{tx} \ac{sdr} gain               & Adjusted for each \ac{tx} position \\
        \ac{RIS} configurations             & 10\,000 per \ac{tx} position \\
        Frames per configuration            & $\approx$ 60 \\
        \ac{RIS} settling time              & \SI{0.3}{\second} (conservative waiting time after reconfiguration) \\
        Transmission window                 & \SI{0.6}{\second} \\
        Silence interval                    & \SI{0.3}{\second} \\
        Environment                         & Controlled indoor with \ac{rf} absorbers \\
        \bottomrule
    \end{tabularx}
\end{table}

\section{Dataset Analysis}
\label{subsec:dataset-analysis}

The objective of this section is to assess how accurately different channel models can reproduce the measured \ac{RIS}-assisted channel response contained in the BRISC dataset. Since the \ac{RIS} employed in the measurement campaign consists of binary-switchable unit cells, each element can assume one of two reflection states, which we denote by $\phi_{\rm on}$ and $\phi_{\rm off}$, corresponding to different complex reflection coefficients that differ in phase and amplitude and depend on the operating frequency.

In an ideal \ac{RIS}-assisted link, the end-to-end channel variation induced by the \ac{RIS} is commonly modeled as a function of the cascaded \ac{tx}--\ac{RIS} and \ac{RIS}--\ac{rx} channels and of the \ac{RIS} configuration vector. However, our measurements show that the channel estimated at the \ac{rx} also contains a configuration-independent component. This component remains present even when the direct \ac{tx}--\ac{rx} \ac{los} is blocked and is mainly due to the \ac{RIS} frame, support structure, residual environmental reflections, and other non-controllable propagation paths. Therefore, in addition to standard \ac{RIS}-dependent models, we also consider models including an additive bias term.

We compare two families of models. The first family consists of linear models, where the \ac{RIS} configuration is mapped to the measured channel through a linear transformation. We consider both a \ac{lm} without bias and a \ac{lmb}, the latter accounting for the configuration-independent channel component observed in the measurements. The second family consists of data-driven nonlinear models, which learn the mapping between \ac{RIS} configurations and channel responses directly from the dataset. In particular, we consider a \ac{nn} and a \ac{raf}.




\subsection{Model Validation}

To validate the models, we used multiple \ac{CSI} samples along with their respective \ac{RIS} configurations to fit/train the different approaches. Let 
\begin{equation}
    \bm{g}_{k}(p) \in \mathbb{C}^{N} 
\end{equation}
denote the \ac{tx}--\ac{RIS} channel column vector at subcarrier $k$, with $k=1, \ldots, K$, where $N$ is the number of \ac{RIS} elements and $K$ the number of subcarriers, when \ac{tx} is located in position $p \in \mathcal{P} = \{p_i\}$, $i=1, \ldots, 9$. Similarly, let 
\begin{equation}
    \bm{F}_{k}(p) \in \mathbb{C}^{M \times N} 
\end{equation} 
denote the \ac{RIS}--\ac{rx} channel matrix, where $M$ is the number of receiving antennas. The \ac{RIS} configuration is represented by the vector $\bm \phi$, whose entries can take one of two possible values, $\phi_{\rm on}$ or $\phi_{\rm off}$, corresponding to the two binary reflection states of each \ac{RIS} element. For each \ac{tx} position and \ac{RIS} configuration, $T$ CSI estimates are available, since several Wi-Fi frames are received while keeping the \ac{RIS} state fixed. Consequently, for a transmitted pilot symbol $x_k$,  the received signal at subcarrier $k$ at frame $t$ is
\begin{equation}
    \bm{y}^{(t)}_k = \bm{H}(\bm{F}_{k}(p),\bm{g}_{k}(p),\bm{\phi}) x_k + \bm{w}^{(t)}_k \in \mathbb{C}^M \,,
\end{equation}
where $\bm{H}(\cdot)$ is the end-to-end channel at the subcarrier $k$ between input $x_k$ and output $\bm{y}_k^{(t)}$ and $\bm{w}_k^{(t)} \sim \mathcal{CN}(\bm{0}, \sigma_w^2\bm{I})$ is the noise vector at frame $t$. Our goal is to estimate the impact that the \ac{RIS} phases have on such a channel.

Let
\begin{equation}
\begin{aligned}
    \hat{\bm H}^{(t)}(\bm{F}_{k}(p),\bm{g}_{k}(p),\bm{\phi}) &=\frac{x_k^* \bm{y}_k^{(t)}}{\lvert x_k \lvert^2}= \\
    &=\bm{H}(\bm{F}_{k}(p),\bm{g}_{k}(p),\bm{\phi})+ \tilde{\bm{w}}_k^{(t)}\,,
\end{aligned}
\end{equation}
be the \ac{CSI} estimate obtained from the $t$-th received frame,
$t = 1, \ldots, T$, we aim to find which model fits best. Specifically, to reduce the effect of measurement noise, we first average all CSI estimates collected under the same \ac{tx} position $p$ and \ac{RIS} configuration $\bm \phi$
\begin{equation}
    \bar{\bm{\bm H}}_{k}(p,\bm{\phi}) =
    \frac{1}{T}
    \sum_{t = 1}^T
    \hat{\bm H}^{(t)}(\bm{F}_{k}(p),\bm{g}_{k}(p),\bm{\phi})\,.
\end{equation}
This gives an averaged CSI value for each subcarrier. The average wideband channel associated with a \ac{RIS} configuration is then obtained by stacking the averaged CSI values over all subcarriers in a matrix as follows
\begin{equation}
    \bar{\bm{\bm H}}(p,\bm{\phi}) = [\bar{\bm{\bm H}}_{1}(p,\bm{\phi}), \ldots, \bar{\bm{\bm H}}_{K}(p,\bm{\phi}) ]^T \in \mathbb{C}^{K \times M} \,.
\end{equation}
The set of measured \ac{RIS} configurations is partitioned into two disjoint subsets: a training set $\mathcal{R}$ and a testing set $\mathcal{I}$ that contain $N_{\rm R}$ and $N_{\rm T}$ \ac{RIS} configurations each. The channels associated with configurations in $\mathcal{R}$ are used to fit or train the model parameters, whereas the configurations in $\mathcal{I}$ are used only for testing. This split allows us to evaluate the ability of each model to generalize to \ac{RIS} states that were not used during fitting.

Let
    $\hat{\bm H}^{\rm m}(p,\bm{\phi})$,
$\rm m \in \{LM,LMB,NN,RAF\}$, be the channel as inferred from the various channel models considered, and better explained next. The performance metric is the normalized average $L_2$ error $\epsilon$ computed between the predicted and measured channel matrices over the testing dataset, i.e.,
\begin{equation}
    \epsilon =
    \frac{
    \sum_{k \in [K],\bm{\phi}\in\mathcal{I}}
    \left\|
   \hat{\bm H}^{\rm m}_{k}(p,\bm{\phi})
    -
    \bar{\bm H}_{k}(p,\bm{\phi})
    \right\|^2
    }
    {
    \sum_{k \in [K],\bm{\phi}\in\mathcal{I}}
    \left\|
    \bar{\bm H}_{k}(p,\bm{\phi})
    \right\|^2
    } .
\end{equation}
In the following, we focus on the CSI measurements obtained with the log-periodic receive antenna, corresponding to the single-antenna case. This choice simplifies the notation without affecting the generality of the conclusions, since the measurements obtained with the dipole antenna lead to the same qualitative behavior.


\paragraph*{Linear Models}
For the $k$-th subcarrier, the \ac{lm} using one antenna is the scalar
\begin{equation}
    H^{\rm LM}_k(\bm{F}_{k}(p),\bm{g}_{k}(p),\bm{\phi}) = \left(
    \left(\bm{F}_{k}(p)\right)^T
    \diamond
    \bm{g}_{k}
    \right)
    \bm{\phi},
\end{equation}
where $\diamond$ denotes the Khatri-Rao product \cite{masood2023inductive}. 
In the \ac{lmb} model a bias $b_k$ is added, leading to
\begin{equation}
    H^{\rm LMB}_k(\bm{F}_{k}(p),\bm{g}_{k}(p),\bm{\phi}) = \left(
    \left(\bm{F}_{k}(p)\right)^T
    \diamond
    \bm{g}_{k}
    \right)
    \bm{\phi} + b_k.
\end{equation}
Let  $\bm{\Phi} \in \mathbb{C}^{N \times N_{\rm R}}$ be the \ac{RIS} phases matrix having columns $\bm{\phi}_n \in \mathcal{R}$,  $\bm{E} \in \mathbb{C}^{K \times N}$  a matrix whose rows are the per-subcarrier Khatri-Rao product, i.e., $\bm{E}_k=\left(
    \left(\bm{F}_{k}(p)\right)^T
    \diamond
    \bm{g}_{k}
    \right)$, and $\bm{Y} \in \mathbb{C}^{K \times N_{\rm R}}$ the measured channel matrix, whose columns are $\bm{Y}_{:,n} = [  \bar{\bm{\bm H}}(p,\bm{\phi}_n)]^T$, $\bm{\phi}_n \in \mathcal{R}$. We can write 
\begin{equation}
    \bm{Y} = \bm{E}\bm{\Phi},
\end{equation}
Therefore, the estimation of $\bm{E}$ can be formulated as a Ridge regression problem, i.e.,
\begin{equation}
    \hat{\bm{E}}
    =
    \arg\min_{\bm{E}}
    \left\{
    \left\|
    \bm{Y}-\bm{E}\bm{\Phi}
    \right\|_{F}^{2}
    +
    \lambda
    \left\|
    \bm{E}
    \right\|_{F}^{2}
    \right\}\,,
\end{equation}
which has a closed-form solution 
\begin{equation}
    \hat{\bm{E}}
    =
    \bm{Y}\bm{\Phi}^{H}
    \left(
    \bm{\Phi}\bm{\Phi}^{H}
    +
    \lambda\bm{I}
    \right)^{-1}\,,
\end{equation}
with $\lambda$ the regularization term aiming at improving numerical stability compared with the conventional least-squares solution. For the \ac{lmb}, an additional constant term is introduced by augmenting the \ac{RIS} configuration matrix as
\begin{equation}
    \tilde{\bm{\Phi}}
    =
    \begin{bmatrix}
    \bm{\Phi}\\
    \bm{1}_{N_{\rm R}}^{T}
    \end{bmatrix},
\end{equation}
where the last row accounts for the bias component, which remains constant across the $N_{\rm R}$ \ac{RIS} configurations. Then, the channel model becomes 
\begin{equation}
    \bm{Y}
    =
    \begin{bmatrix}
    \bm{E} & \bm{b}
    \end{bmatrix}
    \tilde{\bm{\Phi}}\,,
\end{equation}
where $\bm{b} \in \mathbb{C}^K$ is the bias vector, and the Ridge regression solution
\begin{equation}
    \begin{bmatrix}
    \hat{\bm{E}} & \hat{\bm{b}}
    \end{bmatrix}
    =
    \bm{Y}
    \tilde{\bm{\Phi}}^{H}
    \left(
    \tilde{\bm{\Phi}}
    \tilde{\bm{\Phi}}^{H}
    +
    \lambda\bm{I}
    \right)^{-1}.
\end{equation}

The estimated Khatri--Rao matrix and bias vector are then extracted as
\begin{equation}
    \hat{\bm{E}}
    =
    \left[
    \bm{Y}
    \tilde{\bm{\Phi}}^{H}
    \left(
    \tilde{\bm{\Phi}}
    \tilde{\bm{\Phi}}^{H}
    +
    \lambda\bm{I}
    \right)^{-1}
    \right]_{:,1:N},
\end{equation}
and
\begin{equation}
    \hat{\bm{b}}
    =
    \left[
    \bm{Y}
    \tilde{\bm{\Phi}}^{H}
    \left(
    \tilde{\bm{\Phi}}
    \tilde{\bm{\Phi}}^{H}
    +
    \lambda\bm{I}
    \right)^{-1}
    \right]_{:,N+1}.
\end{equation}

Once $\hat{\bm{E}}$ has been estimated, the composite \ac{tx}--\ac{RIS}--\ac{rx} channel can be
predicted for any arbitrary \ac{RIS} configuration $\bm{\phi}_{\rm new}$ as
\begin{equation}
     \bm{H}^{\rm LM}(\bm{F}_{k}(p),\bm{g}_{k}(p),\bm{\phi})
    =
    \hat{\bm{E}}\bm{\phi}_{\rm new},
\end{equation}
and 
\begin{equation}
    \bm{H}^{\rm LMB}(\bm{F}_{k}(p),\bm{g}_{k}(p),\bm{\phi})
    =
    \hat{\bm{E}}\bm{\phi}_{\rm new}
    +
    \hat{\bm{b}}
\end{equation}
for the LM and LMB models, respectively.

\paragraph*{Non-linear Models} To capture the non-linear mapping between \ac{RIS} configurations and resulting \ac{CSI}, we employed two state-of-the-art data-driven approaches, namely \ac{nn} and \ac{raf}. We chose a \ac{nn} for its universal approximation capabilities and \ac{raf} as an ensemble learning strategy suitable for regression tasks when dealing with binary inputs, since each element phase profile can be encoded as $1$ if it is \on, $0$ otherwise. 
In formulae, the input feature vector is defined as $\bm{v}(\bm{\phi}) \in \{0,1\}^N$, 
where the $n$-th entry is given by
\begin{equation}
    v_n(\phi_n) = \begin{cases} 1 & \text{if } \phi_n = \phi_\text{on}, \\ 0 & \text{otherwise}. \end{cases}
\end{equation}

For both learning-based approaches, the same training and testing datasets are employed, namely $\mathcal{D}_{\mathrm{train}}$ and $\mathcal{D}_{\mathrm{test}}$. The training dataset $\mathcal{D}_{\mathrm{train}}$ is defined as
\begin{equation}
    \mathcal{D}_{\mathrm{train}}
    =
    \left\{
    \left(
    \bm{v}(\bm{\phi}_n),
    \bar{\bm{H}}(p,\bm{\phi}_n)
    \right)
    :
    \bm{\phi}_n \in \mathcal{R}
    \right\},
\end{equation}
where $\bar{\bm{H}}(p,\bm{\phi}_n)$ is the  measured composite channel corresponding to the input representation of the \ac{RIS} phase configuration $\bm{v}(\bm{\phi}_n)$. Similarly, the test set $\mathcal{D}_{\mathrm{test}}$ is defined as
\begin{equation}
    \mathcal{D}_{\mathrm{test}}
    =
    \left\{
    \left(
    \bm{v}(\bm{\phi}_n),
    \bar{\bm{H}}(p,\bm{\phi}_n)
    \right)
    :
    \bm{\phi}_n \in \mathcal{I}
    \right\}.
\end{equation}

The \ac{nn} learns a non-linear function that maps the binary RIS configuration $\bm{\phi}_n$ to the corresponding measured wideband channel. Denoting by $\bm{w}$ the trainable parameters of the network, the predicted channel is
\begin{equation}
    \hat{\bm{H}}^{\mathrm{NN}}(p,\bm{\phi}_n)
    =
    f_{\bm{w}}
    \left(
    \bm{v}(\bm{\phi}_n)
    \right).
\end{equation}

The network parameters are optimized by minimizing the squared reconstruction error between the predicted and measured averaged channels defined as
\begin{equation}
    \mathcal{L}
    =
    \left\|
    f_{\bm{w}}
    \left(
    \bm{v}(\bm{\phi}_n)
    \right)
    -
    \bar{\bm{H}}(p,\bm{\phi}_n)
    \right\|_2^2 .
\end{equation}

For the \ac{raf} model, the mapping between the \ac{RIS} configuration and the
channel response is learned through an ensemble of decision trees. The predicted channel is obtained as\looseness=-1
\begin{equation}
    \hat{\bm{H}}^{\mathrm{RAF}}(p,\bm{\phi}_n)
    =
    f_{\mathrm{RAF}}
    \left(
    \bm{v}(\bm{\phi}_n)
    \right)\,,
\end{equation}
where $f_{\mathrm{RAF}}(\cdot)$ denotes the Random Forest regression model. The model is trained using the same training dataset as for \ac{nn} by minimizing the squared reconstruction error.\footnote{Due to space constraints, the parameters of both linear and non-linear models are reported in the repository \cite{gitHubBrisc}.}

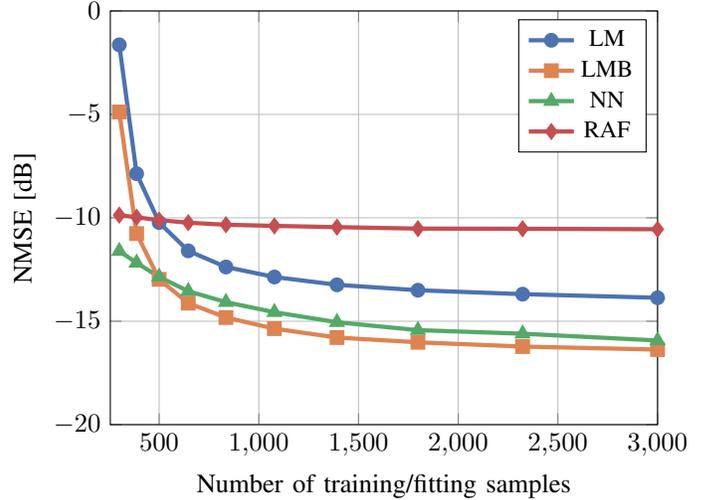
\begin{figure}
    \centering
    \usetikzlibrary{intersections}
\usepgfplotslibrary{fillbetween}

\begin{tikzpicture}

\definecolor{color0}{rgb}{0.298039215686275,0.447058823529412,0.690196078431373}
\definecolor{color1}{rgb}{0.866666666666667,0.517647058823529,0.32156862745098}
\definecolor{color2}{rgb}{0.333333333333333,0.658823529411765,0.407843137254902}
\definecolor{color3}{rgb}{0.768627450980392,0.305882352941176,0.32156862745098}

\begin{axis}[
    width=\linewidth,
    height=0.6\linewidth,
    xlabel={Number of training/fitting samples},
    ylabel ={NMSE [\si{\decibel}]},
    xmin = 255,
    xmax = 3000,
    ymax = 0,
    ymin = -20,
    grid=both,
        legend style={
        at={(0.98,0.98)},
        anchor=north east,
        font=\small 
    }
]


\addplot[
    ultra thick,
    color0,
    mark=*,
] table [
    x index=0,
    y index=1,
    col sep=comma
] {data/mse_samples/linear_overleaf.csv};

\addlegendentry{LM}

\addplot[
    ultra thick,
    color1,
    mark=square*,
] table [
    x index=0,
    y index=1,
    col sep=comma
] {data/mse_samples/linear_bias_overleaf.csv};

\addlegendentry{LMB}

\addplot[
    ultra thick,
    color2,
    mark=triangle*,
] table [
    x index=0,
    y index=1,
    col sep=comma
] {data/mse_samples/nn_overleaf.csv};
\addlegendentry{NN}

\addplot[
    ultra thick,
    color3,
    mark=diamond*,
] table [
    x index=0,
    y index=1,
    col sep=comma
] {data/mse_samples/rf_overleaf.csv};
\addlegendentry{RAF}

\end{axis}
\end{tikzpicture}
    \caption{NMSE for different models as a function of the number of training samples.}
    \Description{Line chart showing NMSE (Normalized Mean Squared Error) for different models as a function of the number of training samples, with NMSE decreasing as the number of training samples increases.}
    \label{fig:learning curves}
\end{figure}

\section{Results}
\label{sec:results}
\Cref{fig:learning curves} shows the reconstruction \ac{NMSE}, averaged across transmitting positions and normalized to the \ac{CSI} power, for different sizes of the fitting/training data. The training dataset was randomly permuted, while testing was performed on the last 2\,000 channels.  All the models achieve less than $-10$ \si{\decibel} error when only 500 samples are used for training, with the non-linear models outperforming the linear models when a small number of training configurations is used.  

\Cref{fig:rec_results} shows the \ac{CSI} reconstructed with \ac{lmb}: the model was fitted on the first $300$ configurations and tested on the subsequent $200$ (301 to 500), without random permutation.
In this setting \ac{lmb} achieves a \ac{NMSE} of \SI{-20.57}{\decibel}, which is lower than that in \Cref{fig:learning curves} for the same number of fitting samples. This suggests that linear models are more effective when we randomly change blocks, rather than changing single elements at random. In fact, changing blocks causes a higher diversity in the resulting channels when compared to configurations where all elements change randomly, and this helps the linear models to better fit the data.

\begin{figure}
  \centering
  \resizebox{\columnwidth}{!}{
  \begin{tikzpicture}
    \begin{groupplot}[
      group style={
        group size=1 by 2,
        vertical sep=12pt
      },
      width=0.8\columnwidth,
      height=0.3\columnwidth,
      scale only axis,
      trim axis left,
      trim axis right,
      xmax=241,
      xmin=0,
    ]

\nextgroupplot[
  ylabel={Magnitude},
  xlabel={},
  xticklabels=\empty,
  xmajorgrids,
  ymajorgrids,
  ylabel style={yshift=10pt},
  ymin=0,
  ymax=10000,
  xtick={0,40,80,120,160,200,240},
  legend style={at={(0.5,1.05)}, anchor=south, legend columns=2}
]

\addlegendimage{black, thick}
\addlegendentry{Predicted}

\addlegendimage{black, dashed, thick}
\addlegendentry{Measured}
    \input{Figures/tikz_folder/reconstructions_magnitude}

    \nextgroupplot[
      xlabel={Subcarrier Index},
      ylabel={Phase [deg]},
      xmajorgrids,
      ymajorgrids,
      ymin=-200,
      ymax=200,
      xtick={0,40,80,120,160,200,240},
    ]
    \input{Figures/tikz_folder/reconstructions_phase}
    \end{groupplot}
  \end{tikzpicture}
}
  \caption{Magnitude (top) and phase (bottom) of the channel frequency response for $4$ representative \ac{RIS} configurations, as reconstructed by the LMB model fit on the first $300$ configurations.}
  \Description{Two stacked line plots sharing the same horizontal axis (subcarrier index, ranging from 0 to 240). The top plot shows the magnitude of the channel response, with solid lines representing predicted values and dashed lines representing measured values, for four different RIS configurations.}
  \label{fig:rec_results}
\end{figure}

\section*{Conclusions}
\label{sec:concs}

This article introduced and released BRISC, a comprehensive meas\-ure\-ment-based dataset for \ac{RIS}-assisted \SI{80}{\mega\hertz} wireless channels in an indoor environment at \SI{5.530}{\giga\hertz}.
By systematically exploring thousands of \ac{RIS} configurations across multiple transmitter locations, BRISC captures the joint impact of \ac{RIS} control, multipath propagation, and spatial geometry under realistic conditions. The dataset was shown to be effective both for calibrating and validating state-of-the-art \ac{RIS} channel models, highlighting the trade-offs between linear and data-driven approaches in terms of accuracy and training requirements. Specifically, a linear model with incorporated bias has been shown to perform similarly, if not even better than a neural network for the channel frequency response reconstruction task. Overall, BRISC constitutes a valuable experimental benchmark for the design, validation, and comparison of signal processing, machine learning, and security techniques for \ac{RIS}-enabled wireless systems.

\balance

\section*{Acknowledgment}
This work was partially supported by Horizon Europe SNS JU through projects 6G-INTENSE (Intent-driven NaTive AI architecturE supporting Compute-Network abstraction and Sensing at the Deep Edge, Grant Agreement N. 101139266), and project PAISES-6G (Pioneering AI-enhanced SEcure 6G Services framework,  Grant Agreement N. 101292896). We thank Prof.~Aydin Sezgin of the Ruhr University Bochum (Germany) for providing the \ac{RIS} used for the experiments.

\bibliographystyle{ACM-Reference-Format}
\bibliography{IEEEabrv, references}

\end{document}